\documentclass[journal=jacsat,manuscript=article]{achemso}
\usepackage{soul}

\usepackage[version=3]{mhchem} % Formula subscripts using \ce{}

\usepackage{amsmath}
\usepackage{amssymb}

\author{Matthijs Berghuis}
\email{a.m.berghuis@tue.nl}
\author{Alexei Halpin}
\author{Quynh Le-Van}
\author{Mohammad Ramezani}
\affiliation[tue]
{Institute for Photonic Integration, Department of Applied Physics, Eindhoven University of Technology,\\
P.O. Box 513, 5600 MB Eindhoven, The Netherlands.\\
Institute for Complex Molecular Systems, Laboratory of Macromolecular and Organic Chemistry, Eindhoven University of Technology,\\
P.O. Box 513, 5600 MB, Eindhoven, The Netherlands.\\
Dutch Institute for Fundamental Energy Research,\\
P.O. Box 6336, 5600 HH Eindhoven, The Netherlands.}
\author{Shaojun Wang}
\affiliation[tue]
{Institute for Photonic Integration, Department of Applied Physics, Eindhoven University of Technology,\\
P.O. Box 513, 5600 MB Eindhoven, The Netherlands.\\
Institute for Complex Molecular Systems, Laboratory of Macromolecular and Organic Chemistry, Eindhoven University of Technology,\\
P.O. Box 513, 5600 MB, Eindhoven, The Netherlands.\\
Dutch Institute for Fundamental Energy Research,\\
P.O. Box 6336, 5600 HH Eindhoven, The Netherlands.}
\author{Shunsuke Murai}
\affiliation[kyoto]
{Department of Material Chemistry, Graduate School of Engineering, Kyoto University, Katsura, Nishikyo, 6158510, Kyoto, Japan}
\author{Jaime G\'omez Rivas}
\email{j.gomez.rivas@tue.nl}
\affiliation[tue]
{Institute for Photonic Integration, Department of Applied Physics, Eindhoven University of Technology,\\
P.O. Box 513, 5600 MB Eindhoven, The Netherlands.\\
Institute for Complex Molecular Systems, Laboratory of Macromolecular and Organic Chemistry, Eindhoven University of Technology,\\
P.O. Box 513, 5600 MB, Eindhoven, The Netherlands.\\
Dutch Institute for Fundamental Energy Research,\\
P.O. Box 6336, 5600 HH Eindhoven, The Netherlands.}

\title[Enhanced Delayed Fluorescence]
  {Enhanced Delayed Fluorescence in Tetracene Crystals by Strong Light-Matter Coupling}


\begin{document}
\begin{abstract}
We demonstrate experimentally an enhanced delayed fluorescence in tetracene single crystals strongly coupled to optical modes in open cavities formed by arrays of plasmonic nanoparticles. Hybridization of singlet excitons with collective plasmonic resonances in the arrays leads to the splitting of the material dispersion into a lower and an upper polariton band. This splitting significantly modifies the dynamics of the photo-excited tetracene crystal, resulting in an increase of the delayed fluorescence by a factor of four. The enhanced delayed fluorescence is attributed to the emergence of an additional radiative decay channel, where the lower polariton band harvests long-lived triplet states. There is also an increase in total emission, which is wavelength dependent, and can be explained by the direct emission from the lower polariton band, the more efficient light out-coupling and the enhancement of the excitation intensity. The observed enhanced fluorescence opens the possibility of efficient radiative triplet harvesting in open optical cavities, to improve the performance of organic light emitting diodes.  
\end{abstract}

\section{Introduction}

Organic semiconductors are important materials for optoelectronic devices such as organic light emitting diodes (OLEDs) and organic photovoltaics (OPV). 
The performance of these devices is determined by properties such as absorption and emission cross section, chemical reactivity and excited state dynamics, arising from the potential energy surface of the molecules.\cite{Kuhn2004,wales_2004} 
Hence, controlling the energy surface can provide a remarkable impact on photo-physical processes involved in these devices.
Tuning of the properties of organic materials is usually done through chemical synthesis. However, changing the molecular composition might also affect the processability and morphology of thin films fabricated from the molecules, which may be detrimental for the performance. 

Recently, an alternative method has emerged to modify the energy surfaces of molecules and to alter their properties without changing the molecular composition: strong light-matter coupling. Theoretical developments continue to uncover the physics underpinning changes to the potential energy surfaces of molecules in this regime \cite{Feist2018,Flick2017,Flick2018}, with tremendous implications for the photophysical properties of organic materials. Strong coupling between photons in optical cavities and excitons in semiconductors results in hybrid quasi-particles called exciton-polaritons. The strong coupling leads to an avoided crossing between the dispersions of cavity photons and excitons at the energy and momentum at which they would overlap. This coupling leads to the splitting in energy of the dispersion into the lower polariton band (LPB) and  the upper polariton band (UPB). The width of the splitting between these bands is determined by the coupling strength and is called the Rabi energy. The properties of these hybrid quasi-particles are a combination of the properties of the two uncoupled states (bare states)\cite{Ebbesen2016a}. This remarkable consequence of strong coupling has led to the possibility of tuning the properties of materials, such as exciton transport~\cite{Schachenmayer2014,Feist2015,Rozenman2018b,Akselrod2010Exciton-excitonMicrocavities}, conductivity~\cite{Orgiu2015}, energy transfer~\cite{Coles2014c,Zhang2014a,Zhong2017,Du2017,Georgiou2018}, or chemical reactivity~\cite{Hutchison2012,Thomas2016,Herrera2016,Galego2016,Luk2017,Peters2017,Munkhbat2018}.

It has been predicted recently that strong coupling can enhance singlet fission in acene molecules\cite{Martinez-Martinez2017}. Singlet fission is the process where one singlet state splits into a triplet pair at approximately half the singlet energy, conserving the total zero spin of the singlet~\cite{Smith2010b}. Singlet fission can lead to a photocurrent quantum efficiency of more than 100\% in a photovoltaic device\cite{Pazos-Outon2017e}. Even higher quantum efficiencies can be achieved by increasing the singlet fission rate\cite{Tayebjee2015,Wu2014c}. Strong light-matter coupling can affect the singlet fission rate, as the energy level of the bright singlet exciton will be modified, while the triplet state remains unaffected. This phenomenon opens the perspective of strong coupling for improving the efficiency of singlet fission based OPV. Also the performance of OLEDs may possibly be enhanced by strong light matter coupling. In the so-called thermally activated delayed fluorescence (TADF) molecules, delayed emission arises from triplets going to the singlet state through thermal activation, in a process known as reverse intersystem crossing (RISC)~\cite{Adachi2014e}. Very recently, a decrease in triplet lifetime in strongly coupled TADF molecule (Erythrosine B) was demonstrated~\cite{Stranius2018} and the inversion of the energy levels of singlet and triplet states was observed in strongly coupled 3DPA3CN\cite{Eizner}. As strong coupling modifies the energy of the coupled transition, it may enhance the rate of RISC, resulting in more efficient emission. Also, enhanced emission originating from fused triplet pair states has been observed for molecules in microcavities, such as TIPS-tetracene\cite{Polak2018ManipulatingMicrocavities}. Common to all these studies is that the cavity used to achieve strong coupling is formed by a pair of mirrors separated by a small distance so that they support a Fabry-P\'erot resonance. However, applications in real devices, such as OPVs or OLEDs, require good accessibility by light from the surrounding medium or efficient outcoupling of light. 

In this manuscript, we demonstrate strong light-matter coupling of excitons in tetracene crystals to an in-plane optical cavity. The cavity consists of silver nanoparticles arranged in a periodic array. This `open' architecture facilitates an efficient excitation of the tetracene and collection of its emission. An advantage of organic single crystals is that we can align the strongest transition dipole of tetracene to the cavity resonance to increase the coupling strength. Strong coupling leads to a Rabi-splitting between the UPB and LPB of 210 meV and we observe a significantly enhanced emission from the lower polariton band, which is a factor of seven higher than the emission of the bare (uncoupled) tetracene at the same wavelength. Time-resolved measurements of the photoluminescence (PL) show an
increase by a factor of almost four of the delayed fluorescence after the normalization of the emission at zero time. Control measurements on a similar sample of tetracene and silver nanoparticles but in the weak coupling regime show the opposite effect, namely, a reduction of the delayed fluorescence with respect to the emission of bare tetracene. Therefore, our measurements illustrate the relevance of strong coupling for the modification of the excited state dynamics of organic semiconductors and the enhancement of the delayed fluorescence, and they open a range of possibilities for improving the emission of OLEDs using strong light-matter coupling of organic semiconductors with in-plane optical cavities.   

\begin{figure}

  \begin{center}
\includegraphics[width=.98\textwidth]{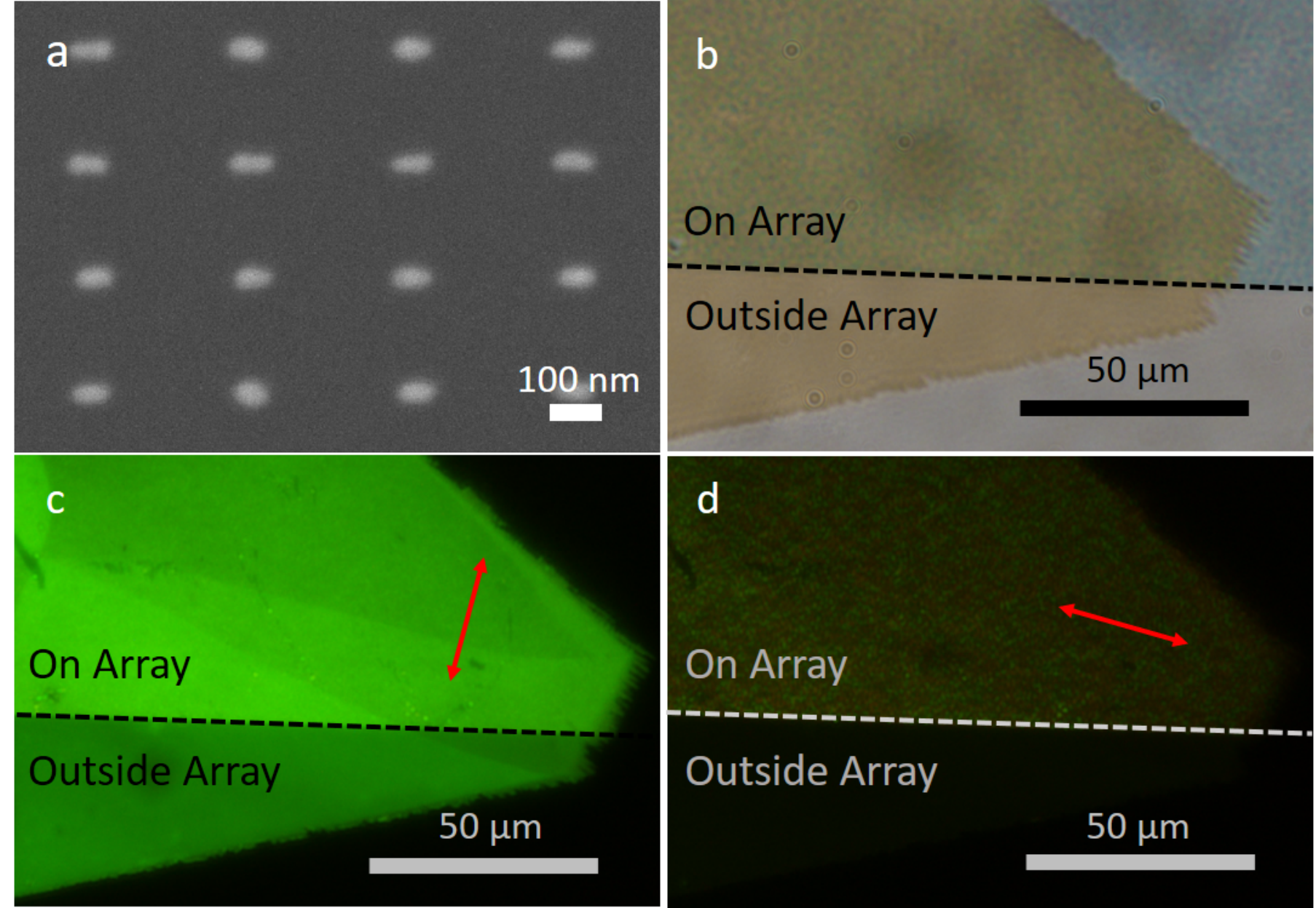}

\end{center}
  \caption{\textbf{a.} SEM image of the array of silver nanoparticles. \textbf{b.} Microscope image of the investigated tetracene crystal in reflection. The crystal is located on the edge of the particle array. \textbf{c.} Fluorescence image of the tetracene crystal obtained with a polarizer parallel to the transition dipole of the lowest optical transition in the crystal. The red arrow indicates the orientation of the polarizer. \textbf{d.} Fluorescence image of the tetracene with a polarizer perpendicular to the lowest optical transition of the crystal.}
  \label{fgr:schematic}
\end{figure}

\section{Plasmonic Nanoparticle Arrays}

Rectangular arrays of silver nanoparticles with dimensions 2.5 $\times$ 2.5 mm$^2$ were fabricated by electron beam lithography on a glass substrate (refractive index $n$=1.51). A 12 nm thick layer of Si$_3$N$_4$ and an 8 nm thick SiO$_2$ layer were deposited on top to prevent oxidation (see Methods and Ref. [\!\!\citenum{Le-Van2019}]). The size of the unit cell is 240 $\times$ 340 nm$^2$ and the nanoparticles have average dimensions of 90 $\times$ 45 $\times$ 40 nm$^3$ (LxWxH). A scanning electron microscope image of the silver nanoparticles is shown in Fig. 1(a). 
The individual nanoparticles support localized surface plasmon resonances (LSPRs), i.e., collective oscillations of the free electrons driven by electromagnetic fields. 
LSPRs confine light to a sub-wavelength scale, leading to large near-field enhancements.~\cite{Muhlschlegel2005} Due to Ohmic losses in the metal and radiation losses to free space, LSPRs are broad and have a relatively low quality factor.~\cite{Wang2006} By carefully arranging the nanoparticles in a periodic array, it is possible to significantly reduce the losses~\cite{Zou2004}. This reduction originates from the enhanced radiative coupling of LSPRs in the array through in-plane diffraction orders known as Rayleigh anomalies (RAs). The resulting modes are known as surface lattice resonances (SLRs), extending over several unit cells of the array. SLRs define a cavity mode with a much longer lifetime than LSPRs due to destructive interference of light scattered to the far field and the redistribution of the near field farther from the metal surface, which results in a reduction of the Ohmic losses~\cite{Zou2004,Guo2015b}. 
The energy and quality factor of the SLRs can be tuned by changing the size of the particles, the lattice dimensions and the number of particles in the unit cell.\cite{Schaafsma2016,Baur2018a,Guo2017b}

We have characterized the SLRs of the nanoparticle array by covering it with a 200 nm thick layer of polystyrene (PS) with a refractive index $n$=1.59. This refractive index is similar to that of the tetracene crystal at frequencies other than the exciton frequency~\cite{Tavazzi2008a}. The similar dielectric environment ensures that the extinction measurements of this sample, given by 1-transmittance, can be used later as a reference for the measurements of the array covered with tetracene. The extinction by the array of nanoparticles covered with PS is measured using a collimated white light source polarized along the short axis of the nanoparticles, while rotating the sample (see Methods). In this way, we obtain the extinction as a function of energy and wave vector of the incident beam parallel to the surface, $k_p= \frac{\omega}{c_0}\sin(\theta)$, where $c_0$ is the speed of light in vacuum, $\omega$ the angular frequency of the light, and $\theta$ the angle of incidence. This extinction map is shown in Fig. 2(a), for photon energies from 2.2 to 2.6 eV and $k_p$ from 0 to -8 rad/$\mu$m. Only negative angles are plotted in this figure because of the symmetry in the extinction around $k_p=0$ rad/$\mu$m. We can identify a parabolic-shaped band of increased extinction with a minimum energy of 2.35 eV. This sharp band corresponds to the SLRs that results from the coupling between the LSPRs along the short nanoparticle axis (visible at 2.7 eV in the blue dash-dotted curve in Fig. 2(e)) and the RAs resulting from the (0, $\pm$1) diffractive orders. 

To obtain the dispersion of the SLRs, we fit the extinction measurements with a coupled harmonic oscillator model~\cite{Ramezani2018}, in which one oscillator has the energy of the LSPR and the other corresponds to the RAs. The fit is plotted with the red dashed curve in Fig. 2(a).
In the extinction spectrum of the bare array, there is a second band visible just above the main resonance at 2.46 eV. This band is attributed to a quasi-guided mode in the PS layer that results from the grating coupling assisted by the array~\cite{Murai2013,Christ2003}.
From the characterization in Fig 2 (a), it is clear that the array is really `open': the array is transparent for most energies and light is coupled into the cavity efficiently at the indicated dispersion of the SLR. In Fig. S4 of the supplemental information (SI), we display the distribution of the electric fields at the energy of the SLR obtained from Finite Difference in Time Domain (FDTD) simulations using a commercial software (Lumerical). These simulations show enhanced fields by a factor of more than 10 in the PS layer and up to 150 nm above the substrate under normal incidence plane wave illumination. The open characteristics and the extended electromagnetic fields associated to SLRs make nanoparticle arrays ideally suitable for light emission applications.

\begin{figure}

  \begin{center}
\includegraphics[width=.98\textwidth]{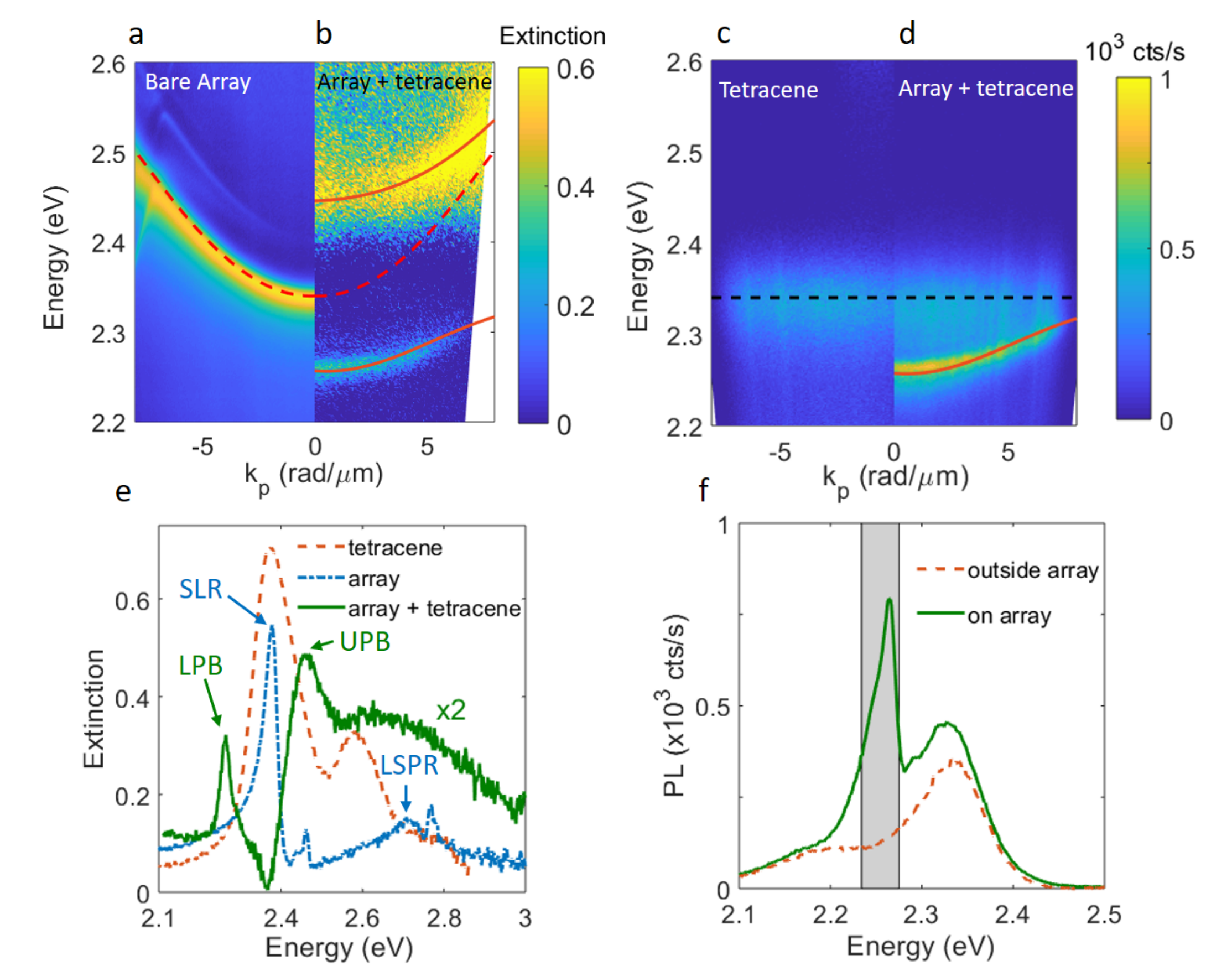}

\end{center}
  \caption{\textbf{a.} Optical extinction map of the nanoparticle array covered with polystyrene (n=1.59). The sample is illuminated with white light polarized along the short axis of the silver particles. The parabolic band of high extinction corresponds to the (0, $\pm 1$) SLRs. The red dashed curve is a fit of this mode. \textbf{b} Extinction map of the particle array covered with the tetracene crystal and referenced to the tetracene crystal outside the array. The extinction is multiplied by a factor two to increase the contrast in the image. The UPB and LPB are fitted with a coupled oscillator model and shown with the red solid curves. The dashed black line at 2.38 eV represents the energy of the S$_0$ $\rightarrow$ S$_1$ transition in tetracene. \textbf{c.} Fluorescence map of the tetracene crystal outside the array. \textbf{d.} Fluorescence  map of the tetracene crystal on top of the nanoparticle array. The coupled oscillator model of Fig. 2(a) is plotted to show that the emission follows the LPB. \textbf{e.} Extinction spectra of the bare tetracene (orange dashed curve), of the array with polystyrene (blue dash dotted curve) and the array with tetracene (green solid curve), showing the Rabi-splitting between LPB and UPB of 210 meV. All spectra are taken at k$_p$ = 4 rad/$\mu$m. \textbf{f.} Fluorescence spectra of tetracene on (green solid curve) and outside the array (orange dashed curve) at k$_p$ = 0 rad/$\mu$m. The semi-transparant gray box indicates the transmission window of the 550 nm bandpass filter used for the time resolved measurements.}
  \label{fgr:extinction}
\end{figure}

\section{Modification of the Absorption and Emission of Tetracene Crystals coupled to Nanoparticle Arrays.}

After the characterization of the bare array of nanoparticles, we removed the PS layer and deposited tetracene single crystals on top of the array. Tetracene has been thoroughly investigated for its semiconducting properties and singlet fission~\cite{Hofberger1975,Vaubel1969,Merrifield1969}. The crystals have a herringbone structure with two molecules per unit cell\cite{Davydov1964TheExcitons}. The transition dipole moment of the lowest optical transition lies almost, but not exactly, in the (001) plane, which is parallel to the substrates where crystals are grown~\cite{Tavazzi2008a} (see Fig. S2 in the SI for X-ray diffraction (XRD) measurements). By drop casting a saturated solution of tetracene in toluene in a nitrogen glovebox~\cite{Burdett2012a} (see Methods), thin tetracene single crystals were formed with a size distribution in the range of 20-300 $\mu$m and 100-400 nm thickness. As the photophysical properties between different tetracene crystals may vary due to different thickness and size, we chose to investigate the crystal shown in Fig. 1(b-d) because it is located at the edge of the nanoparticle array. In this way, the properties of the same crystal could be compared on and outside the array. This crystal has a thickness of 140$\pm 50$ nm (see Methods), and a size of $\sim 250 \times 200\;\mu$m$^2$. Figure 1(b) shows the crystal in reflection under white light illumination. The black dashed line in Fig. 1(b) indicates the edge of the nanoparticle array. There is only a small contrast between the reflection inside and outside the array. The fluorescence of the tetracene crystal, diplayed in Figs. 1(c,d), shows a higher contrast due to the stronger emission from the array. In Fig. 1(c), the fluorescence of the crystal is imaged by illuminating the sample with a white light source in combination with a 400 nm short pass filter and with a polarizing filter parallel (red arrow) to the lowest optical dipole transition of the tetracene crystal, while in Fig. 1(d) the polarizing filter is perpendicular (red arrow) to the lowest optical dipole transition. We conclude from these images that the orientation of the dipole moment of the lowest optical transition is at an angle of 10$^\circ$ with respect to the dipole moment of the LSPR of the short axis of the silver nanoparticles (note that this axis is perpendicular to the edge of the array, visible in Figs. 1(c,d)). This orientation leads to a strong interaction between tetracene excitons and SLRs arising from the diffractive coupling of LSPRs with their dipole moments along the short nanoparticle axis.~\cite{Hertzog2017}.

The result of the coupling between excitons in tetracene and the SLRs in the array is visible in the extinction map of Fig. 2(b), where the extinction is measured in a Fourier microscope (see Methods). For a direct comparison with the dispersion of the bare nanoparticle array, we plot in this figure only the dispersion for positive wave vectors (the dispersion for negative wave vectors is similar because of the symmetry of the array). The dashed black line and dashed red curve in this figure indicate the dispersion of the lowest exciton transition in tetracene and the SLRs of the bare array for the uncoupled case. 

A characteristic of strong coupling is the avoided crossing between the bare states and the formation of the hybrid states given by the LPB and UPB. In the case of the investigated nanoparticle array, the  avoided crossing is apparent at $k_p=4$ rad/$\mu$m and $E=2.38$ eV (See Fig. 2(b)). The LPB and UPB can be fitted to the eigenfrequencies of the Hamiltoninan described by a coupled harmonic oscillator model~\cite{Rodriguez2013a},
\begin{equation} \label{coupledOscillator}
   H=
  \left[ {\begin{array}{cc}
   E_{SLR}-i\gamma_{SLR} & g \\ 
   g & E_{exc}-i\gamma_{exc} \\
  \end{array} } \right] \; ,
\end{equation}
\newline
where E$_{SLR}$ is the angle dependent energy of the degenerate SLRs, shown with the red dashed curve in Figs. 2(a) and (b), E$_{exc}$ is the energy of the tetracene S$_1$ exciton (2.38 eV), $\gamma_{SLR}$ and $\gamma_{exc}$ are the damping rates of the SLRs and the exciton, respectively, estimated from the full width at half maxima of the extinction spectra of the bare modes, and $g$ is the coupling strength from which the Rabi-splitting ($\Omega = 2g$) can be calculated. The result of this model is plotted with the red solid curves in Fig. 2(b), and it fits well to the experimental data. The Rabi-splitting obtained from the fit ($\Omega = 210$ meV) is larger than the damping rates of the SLRs ($\gamma_{SLR}= 30 $ meV) and the linewidth of tetracene ($\gamma_{exc}=160$ meV), confirming that the system is in the strong coupling regime. 

To better visualize the Rabi-splitting, we plot in Fig. 2(e) the extinction spectra for bare tetracene (outside the array), for the bare nanoparticle array and for the array coupled to tetracene, measured at $k_p=4\; $rad/$\mu$m. The dashed orange curve shows the extinction of the tetracene crystal, where the maximum extinction at 2.38 eV corresponds to the lowest optical exciton transition. Higher vibronic transitions are visible at $\sim 2.58$ eV and $\sim 2.76$ eV. The maximum extinction of tetracene overlaps in energy with the maximum extinction of the bare array for this wave vector (blue curve in Fig. 2(e)), corresponding to the SLR. The solid green curve in Fig. 2(e) represents the extinction of the coupled system, referenced to the extinction of tetracene. We observe the splitting of the exciton energy into the LPB and the UPB at 2.26 eV and 2.47 eV, respectively. The broad peak around to 2.7 eV, visible in the extinction of the array with tetracene and the array with PS, corresponds to the LSPRs of the individual silver nanoparticles. 

We have measured the fluorescence of the tetracene crystal on and outside the nanoparticle array using a pulsed laser diode with $\lambda=375$ nm (3.3 eV) and a repetition rate of 2.5 MHz to excite the organic crystal off resonance. Note the bare array (without tetracene) is formed by Ag nanoparticles and it does not exhibit fluorescence. Figures 2(c) and (d) show a map of the fluorescence of bare tetracene and the fluorescence of tetracene on the array, respectively. Similar to the extinction measurements, the emission is symmetric around $k_p=0$ rad/$\mu$m, and we only plot this emission for half the space. The emission outside the array is mainly originating from the lowest exciton transition, indicated by the black dashed line at 2.35 eV on Figs. 2 (c,d). The intensity of the emission at this energy on the array is almost unchanged, which indicates that there is a significant fraction of dark (uncoupled) excitons. However, there is a very prominent peak visible at 2.26 eV for $k_p=0$ rad/$\mu$m, corresponding to the LPB. At this frequency, the emission is seven-fold more intense than the emission at the same frequency of the tetracene outside the array. This emission follows the dispersion of the LPB as it was determined from the coupled oscillator Hamiltonian, and plotted with the lowest red curve in Fig. 2(d). There is no emission from the UPB, in agreement with previous measurements of organic strongly coupled systems, and explained by the fast relaxation to dark excitons and to the LPB~\cite{George2015}. For an easier comparison, we have plotted the fluorescence emission spectra at $k_p=0$ rad/$\mu$m on and outside the array in Fig. 2(f) with the solid green and red dotted curves, respectively.

\begin{figure}
  \begin{center}
\includegraphics[width=.49\textwidth]{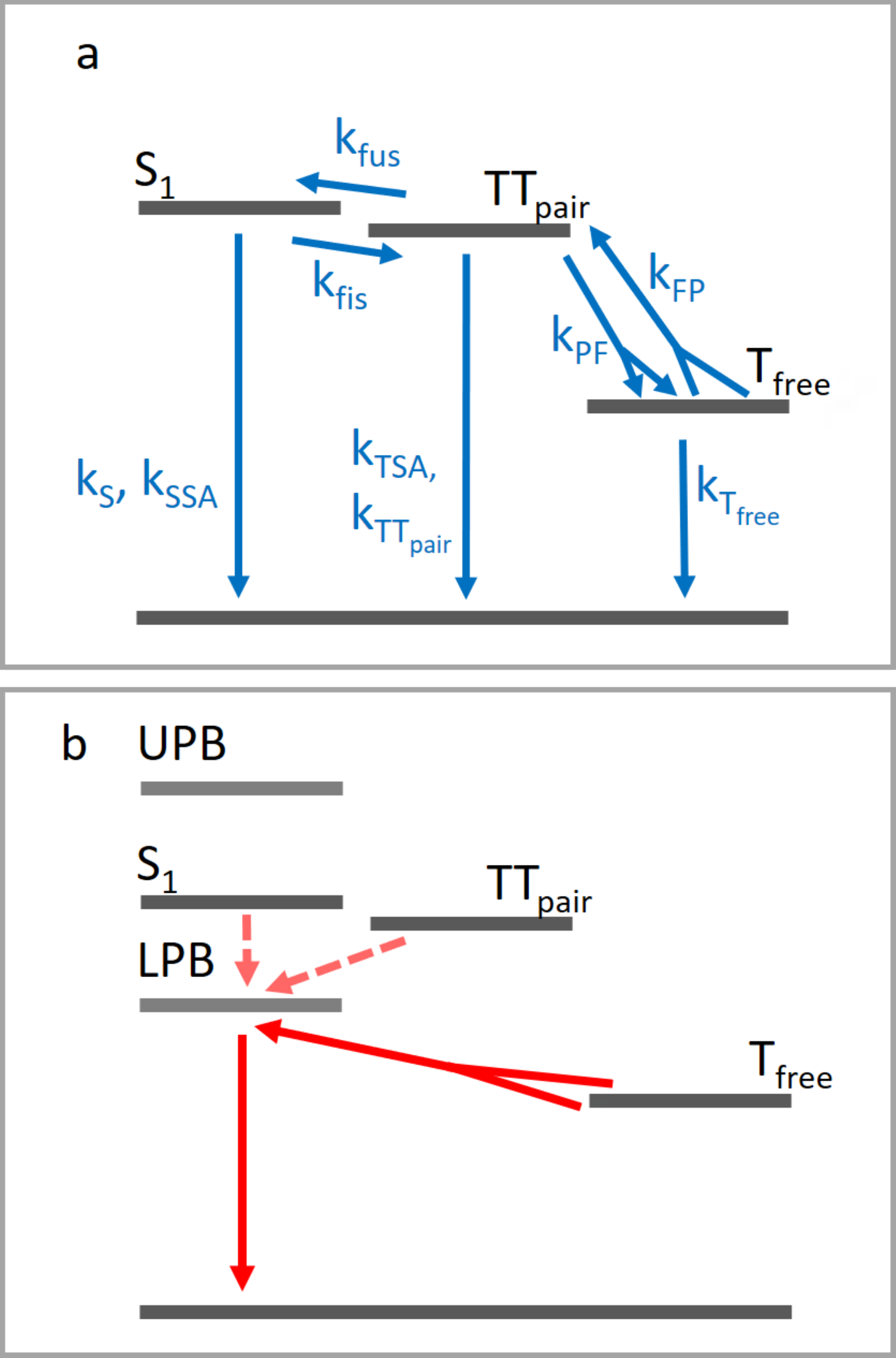}

\end{center}
  \caption{\textbf{a.} Jablonski diagram showing the main transitions in the tetracene crystal. The TT$_{pair}$ state represents all triplet pair states and T$_{free}$ is the free triplet state. Rates between different populations are indicated with the blue arrows. \textbf{b.} Jablonski diagram of a tetracene crystal strongly coupled to the array. Only the new possible decay channels originating from strong coupling, have been indicated with the red and red dashed arrows. The decay channel of free triplet states via the LPB is indicated with the thicker arrows. This relaxation pathway provides the largest modification in the delayed fluorescence.}
  \label{fgr:A}
\end{figure}

\section{Excited State Dynamics of Bare and Strongly Coupled Tetracene Crystals}
In what follows, we investigate how strong coupling affects the photo-physics of single crystal tetracene. Figure 3(a) shows a schematic diagram with the most important excited state dynamics of the tetracene crystal, which can be summarized by the reaction equation 
\begin{equation} \label{simpleRate}
S_0+S_1\rightleftarrows TT_{pair}\rightleftarrows 2\; T_{free}\,.
\end{equation}
The excited singlet ($S_1$) together with a singlet ground state ($S_0$) can turn into a pair of triplet states ($TT_{pair}$). The $TT_{pair}$ can separate into two free triplets ($T_{free}$). The process from the singlet state to the triplet pair state is spin allowed as the total spin of the two bound triplets adds up to zero. The singlet energy in tetracene crystals is approximately 180 meV lower than twice the energy of a free triplet\cite{Tomkiewicz1971}. Despite this apparent energy mismatch, singlet fission to a triplet pair state proceeds at a rate of $\sim$8.3 ns$^{-1}$ \cite{Wan2015a}.
The opposite process to singlet fission is triplet fusion, where the triplet pair fuses to form a singlet state (from $TT_{pair}$ to $S_0 +S_1$ in Eq. \eqref{simpleRate}). Triplet fusion results in a delayed component of the tetracene PL, that dominates the decay at intermediate times, roughly between 1 and 5 ns. Besides fusion to a singlet state, the triplet pair can separate into free triplets that can diffuse through the crystal\cite{Stern2015a,Mauck2016,Burdett2011b,Chan2012a,Smith2013}. When two free triplets meet, they may form a triplet pair and, if the zero spin criterion is fulfilled, they can annihilate to form a singlet exciton\cite{Tayebjee2013a,Smith2010b}. In this manuscript, we refer to the process of the annihilation of two free triplets to a singlet exciton, as free triplet annihilation, while the annihilation of a triplet pair is referred to as triplet fusion. 
As the annihilation of two free triplets to a singlet depends on the probability of two triplets to interact, this rate scales quadratically with the number of excitons and therefore with the excitation intensity.~\cite{Bayliss2014}
Free triplet annihilation is the main origin of the delayed fluorescence from tetracene on time scales longer than 10 ns~\cite{Burdett2010a,Yong2017e,Akselrod2014h,Wan2015a}.

By strong coupling of the singlet exciton transition in tetracene to a plasmonic cavity, we form the LPB with partial singlet character, while the triplet energy remains the same. A schematic representation of the effect of strong coupling on the tetracene energy levels is given in Fig 3(b). We hypothesize that strong coupling may lead to an additional decay channel of triplet-triplet annihilation to the LPB.  Because of this additional channel and the brightness of the LPB in Fig. 2(d), strong light-matter coupling may lead to an enhanced delayed fluorescence \cite{Polak2018ManipulatingMicrocavities}. To investigate if strong coupling indeed affects the excited state dynamics of the system, we have analyzed the time-resolved fluorescence using time-correlated single photon counting. For clarity, these measurements have been done by collecting mainly the fluorescence from the LPB using a band pass filter centered at $\lambda=550$ and with a bandwidth of 10 nm. The detected wavelength range is indicated by the gray box in Fig. 2(f). Similar measurements have been done with a $\lambda=530 \pm 5 nm$ band pass filter to collect the emission from the dark (uncoupled) excitons. These measurements, which are not discussed in the manuscript but can be found in the SI (Fig. S8), conclude that dark excitons inherit the properties of the LPB, as has been recently predicted.~\cite{Gonzalez-Ballestero2016}

The time resolved fluorescence (TRPL) decay of the crystal outside the array is shown by the orange solid curve in Fig. 4(a). The initial exciton density is estimated from the absorbed laser power and the size of the excitation spot to be $8 \times 10^{17}$ cm$^{-3}$. Also the instrument response function (IRF), with a full width at half maximum of $\tau_{IRF}=450$ ps, is shown in Fig. 4(a) with the purple dash-dotted curve. In the decay of the bare tetracene measured outside the array, we observe three regions: a very fast decay that is convoluted with the IRF of the setup (see inset of Fig. 4 (a)), an intermediate decay, i.e. from $\sim$1 until $\sim$5 ns, and a slow decay component from $\sim$10 ns after excitation. The TRPL measurements of tetracene on the array (green dashed curve in Fig. 4 (a)) display the same three regions. There are however two striking differences: at $t=0$, the emission intensity of tetracene on the array is already a factor 3 higher than outside the array (green dashed curve in the inset of Fig. 4(a)). This higher prompt fluorescence intensity can be attributed to a faster singlet (radiative) decay to the ground state via the lower polariton band, to pump enhancement, or to a more efficient light out-coupling. However, we focus here on the change in decay rates rather than the absolute fluorescence intensity. 

If we follow the decay of the tetracene emission on and outside the array, we see that these curves are parallel during the first 5 ns (See inset of Fig. 4 (a)). After $\sim$5 ns, the contribution of the delayed emission on the array becomes larger compared to the bare tetracene.

To show the difference in the excited state dynamics on and outside the array more clearly, the decay curves have been normalized to the same maximum at t=0, and are plotted in Fig. 4(b). The red curve represents the time-resolved fluorescence outside the array, while the green curve is the measurement of the strongly coupled tetracene under the same conditions. When we compare these decays, we see that they nearly overlap during the first $\sim$5 ns, after which they start to diverge quickly over a time period of $\sim$50 ns. To better describe the dynamics of the excited states of the tetracene crystal and to understand what may be the cause of the difference in TRPL, we have used a kinetic model to describe the time evolution of the different states~\cite{Burdett2010a,Burdett2013a,Wan2015a,Wilson2013c,Birech2014a}. To simplify the model and limit the number of fitting parameters, we did not consider the different spin states of the triplet pair as separate states,\cite{Tayebjee2017d,Yong2017e,Weiss2017b} but they are all included in the triplet pair state. The model is given by the following rate equations:

\begin{equation} \label{popS}
\frac{dN_S}{dt} = -(k_S + k_{fis})N_S-k_{SSA}N_{S}^2+k_{fus}N_{TTpair},
\end{equation}
\begin{equation} \label{popTT}
\frac{dN_{TTpair}}{dt} = -(k_{TTpair} + k_{fus} + k_{PF})N_{TTpair}+k_{fis}N_{S}+k_{FP}N_{Tfree}^2,
\end{equation}
\begin{equation} \label{popTfree}
\frac{dN_{Tfree}}{dt} = -k_{Tfree}N_{Tfree}-2k_{FP}N_{Tfree}^2+2k_{PF}N_{TTpair}.\newline
\end{equation}

In these equations, N$_S$, N$_{TTpair}$ and N$_{Tfree}$ represent the populations of the singlet state, the triplet pair and the free triplets, respectively.  The radiative and non-radiative decay rates of the singlet, triplet pair, and free triplet state are given by k$_S$, k$_{TTpair}$ and k$_{Tfree}$. k$_{SSA}$ is the rate of singlet-singlet annihilation, which is a second order process, as is triplet singlet annihilation,  k$_{TSA}$. k$_{fis}$ and k$_{fus}$ are the rate of singlet fission into a triplet pair state and the rate of fusion from a triplet pair state into a singlet state, respectively. Because triplet fusion is the process in which one bound triplet pair transfers into one singlet state, this process is linear with the triplet pair density. As two free triplets can form one triplet pair state, k$_{PF}$ and k$_{FP}$ have a prefactor of two in Eq. \eqref{popTfree}, and the decay rate of free triplets to the triplet pair depends quadratically on the free triplet concentration\cite{Tayebjee2013a,Burdett2010a,Yong2017e}. The time step used for the numerical calculation of Eqs. 3-5 is 10 ps. We assume that right after excitation all the excited states are in the S$_1$ state, thus the TT$_{pair}$ and $T_{free}$ are not excited directly. To limit the amount of fitting parameters and because some decay rates are faster than the IRF of our setup, we fixed most of the rates to the values given in Ref. [\!\!\citenum{Wan2015a}]. The only free parameters of the fit were the decay rate of the triplet state, the triplet pair state, and the initial population of the $S_1$ states. We left these values free since in Ref.[\!\!\citenum{Wan2015a}] the focus is on the decay during the first 10 ns, while we are interested in times up to 250 ns. A particle swarm algorithm is used to fit the measurements in the interval 2-250 ns after excitation.~\cite{494215} The black dashed curve in Fig. 4 (c) shows the fit to the TRPL data measured outside the array. The fitting parameters are given in Table 1 and the residuals are plotted in the SI (Fig. S7). The fit of the kinetic model is in good agreement with the measured data for most of the fitted range. There is a small discrepancy between the fit and the data around 10 ns, resulting from the approximations made in the model: In particular, we do not consider the spin evolution of the triplet pair states, and disregard emission from any states other than the singlet state. Nevertheless, the kinetic model fits the largest fraction of the decay curve excellently. Therefore, we can conclude that the kinetic model describes the main processes in the crystal reasonably well, and gives reliable values for the populations of the different excited states over time.

The populations of the different excited states are plotted in Fig. 4 (c) for an initial excitation density of $8.1 \times 10^{17}$ cm$^{-3}$, normalized to the S$_1$ population at t=0. During the first $\sim$200 ps, the $S_1$ state is the dominant population, plotted with the blue curve in Fig. 4 (c) (see also the inset). However, efficient singlet fission results in a rapid population of the TT$_{pair}$ state, shown with the green dashed curve. These triplet pairs split into free triplets (orange dash-dotted curve) and free triplets become the most populous species in the system after $\sim$1 ns. At $\sim$10 ns there is a kink in the singlet population caused by the fact that from that time the S$_1$ decay dynamics become dominated by the free triplet reservoir. The reason why it takes a few ns after the triplet reservoir is the most populated state until it starts to dominate the singlet fluorescence decay, is because the transition rates from the triplet to S$_1$ are slower than from the triplet pair states to S$_1$. If we compare Fig. 4 (c) with Figs. 4 (a) and (b), we conclude that the fastest decay component of the tetracene fluorescence corresponds to the ultrafast singlet fission from S$_1$ to the TT$_{pair}$ state. Then the S$_1$ state is repopulated by TT$_{pair}$ states, while at the same time the TT$_{pair}$ dissociates into free triplets. Free triplets act as an exciton reservoir that repopulates the singlet state through free triplet annihilation from times longer than $\sim$10 ns after excitation.

 \begin{figure}
  \begin{center}
\includegraphics[width=.48\textwidth]{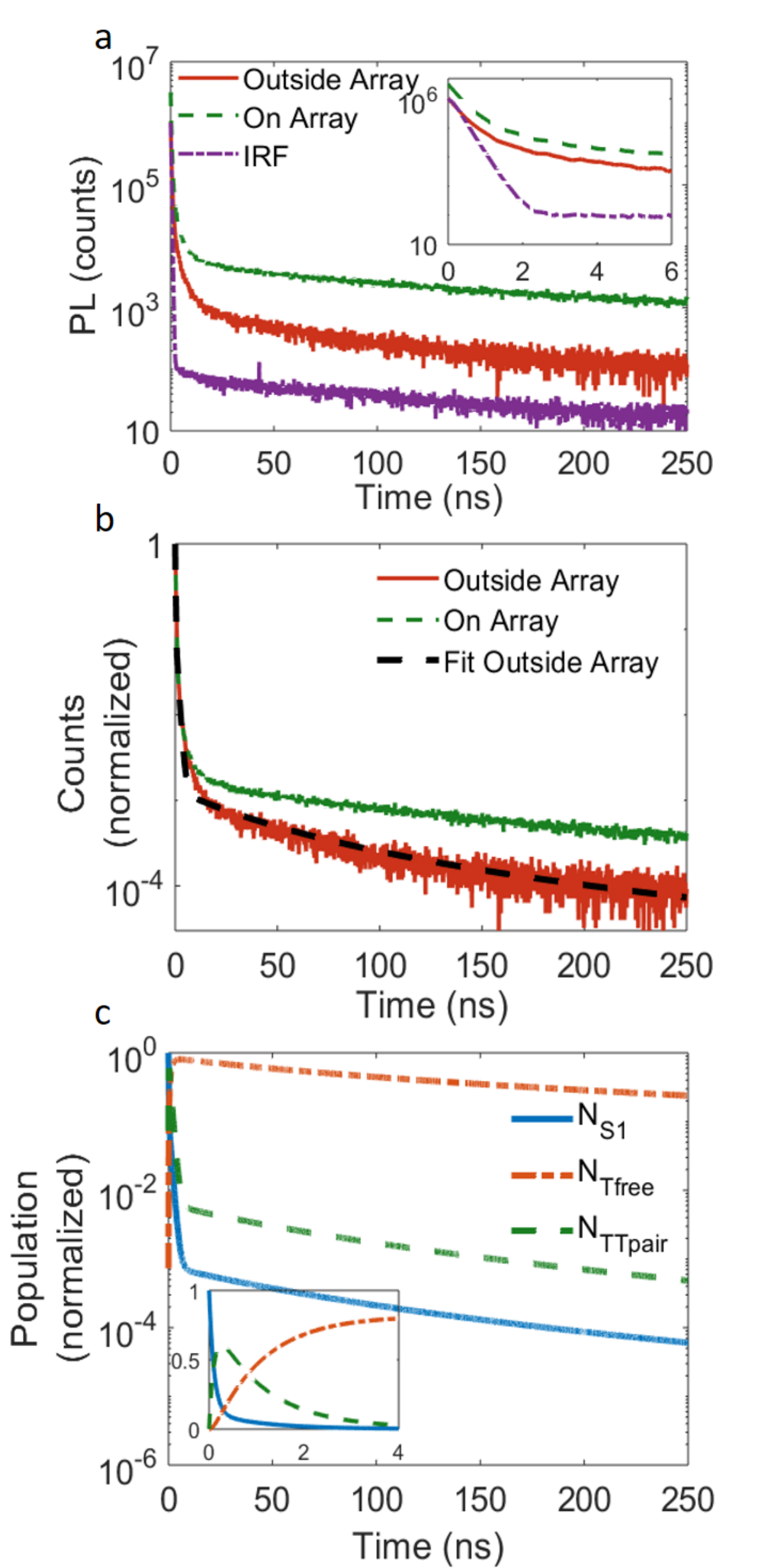}

\end{center}
  \caption{time-resolved PL (TRPL) of tetracene. \textbf{a.} TRPL measurements on (green dashed curve) and outside (red solid curve) the array are shown. The initial singlet exciton density is $N_S \sim 8.1 \times 10^{17}$ cm$^{-3}$. The instrument response function (IRF), shown with the purple dash-dotted curve, has a FWHM of 450 ps and is normalized to the decay outside the array. The inset shows the same three curves, but for the first 6 ns. \textbf{b.} TRPL for tetracene on (green curve) and outside (red curve) the array normalized at t=0. The fit to the TRPL outside the array is shown with the black dashed curve. The fitting parameters can be found in table~\ref{tbl:k}. \textbf{c.} $S_1$, $T_{pair}$ and $T_{free}$ populations as a function of time as obtained from the kinetic model for the initial excitation density of  $N_{S}=3=8.1 \times 10^{17}$ cm$^{-3}$. The inset shows the rapid decay of the singlets to the triplet pair states in $\sim200$ ps and the build up of the free triplets in the first nanoseconds.}
  \label{fgr:TRPL}
\end{figure}

\begin{table}
  \caption{Rates describing the dynamics of excitons in single crystal tetracene. The initial singlet exciton population is estimated from the excitation intensity and the absorption by tetracene, and subsequently fitted to the experiments. To minimize the number of fitting parameters, the only free parameters are $k_{TTpair}$, $k_{Tfree}$ and $N_{S0}$. All the other rates have been taken from literature.\cite{Wan2015a}. The results of the fit with these parameters are shown with the black dashed curve in Fig 4(b).}
  \label{tbl:k}
  \begin{tabular}{lll}
    \hline
    symbol  & rates  & process  \\
    \hline
    $k_S$   &   $0.08$ ns$^{-1}$ & decay of singlets \\
    $k_{TTpair}$ &  $0.44$ ns$^{-1}$  &decay of triplet pair \\
    $k_{Tfree}$ & $0.05\; \mu$s$^{-1}$  & decay of free triplet\\
    $k_{SSA}$ & $2.0 \times 10^{-18}$ ns$^{-1}$cm$^{-3}$  & singlet-singlet annihilation\\
    $k_{TSA}$ & $2.5 \times 10^{-18}$ ns$^{-1}$cm$^{-3}$  & triplet-singlet annihilation\\
    $k_{fis}$ & 8 ns$^{-1}$  & $S_0 + S_1 \rightarrow TT_{pair}$\\
    $k_{fus}$ & $1$ ns$^{-1}$ & $TT_{pair} \rightarrow S_0 + S_1$\\
    $k_{PF}$ & $0.5$ ns$^{-1}$ & $TT_{pair} \rightarrow$ 2 $T_{free}$\\
    $k_{FP}$ & $1.8 \times 10^{-20}$ ns$^{-1}$cm$^{-3}$
      & $2 \space T_{free} \rightarrow TT_{pair}$\\
     $N_{S0}$   &  $4.6 \times 10^{17}$ cm$^{-3}$  & singlet population at t=0 \\
    \hline
  \end{tabular}
\end{table}

We note that strong coupling modifies the energy landscape, creating new energy bands. Adding new states to the kinetic model also increases the number of fitting parameters, reducing the reliability of the fits. Although with the kinetic model we cannot describe the effects of strong coupling quantitatively, there is clearly a large effect of strong coupling on the delayed PL. This difference becomes even more evident when we plot the ratio between the fluorescence on and outside the array as it is done with the red open circles in Figs. 5(a) and (b). This curve is obtained by dividing the normalized decay on the array by the normalized decay outside the array, and taking the average of this ratio in steps of 8 ns. The error bars in Figs. 5(a) and (b) represent the standard deviation of this average. The ratio between the fluorescence on and outside the array remains one for $\sim$5 ns (see red circles in inset Fig. 5(b)), then rapidly increases to a factor of $\sim$ 2.5 at 50 ns after excitation, and continues to increase for 200 ns but at slower rate. To find a qualitative explanation for this ratio as a function of time, we look at the different excited states in tetracene, obtained from the kinetic model in Fig. 4 (c). From this model, we know that during the first $\sim$5 ns the excited state dynamics is dominated by the S$_1$ and ${\rm TT}_{pair}$ states, consecutively. Since at this time interval the ratio between fluorescence on and outside the array is constant and nearly one (see red open circles in the inset of Fig. 5(b)), we can conclude that the S$_1$ and ${\rm TT}_{pair}$ states do not play a significant role in the observed enhanced delayed emission. The rapid increase of this ratio between 5 and 50 ns coincides with the time that the triplet states become the most dominant species in the system ($\sim$10 ns). This behavior strongly suggests that the change in the time-resolved PL between the bare tetracene and the strongly coupled tetracene is related to triplet states. Indeed, the enhanced delayed emission due to the triplet states could be explained by the formation of an additional radiative channel from the triplet states to the LPB in the strong coupling regime. A more efficient harvesting of the triplet states by the LPB could be expected by considering that this pathway is energetically more favourable since the LPB is at lower energy than the S$_1$ state. However, in this case we should also expect a faster depletion of the triplet reservoir, leading to a faster decay of the triplet states, which is not what we see in Figs. 4 and 5. This apparent contradiction could be explained if we consider that once the LPB has ‘harvested’ the triplets, the emission is more efficient than for the bare tetracene. This would imply that the number of triplet states and the number of harvested triplets are almost the same for the uncoupled and the strongly coupled tetracene, but the in the last case the harvested triplets by the LPB decay radiatively, emitting more efficiently than the triplets harvested by $S_1$ state in the uncoupled case. A more plausible explanation could be that strong coupling modifies more than only the decay rates towards the LPB - it may change any rate between the LPB and UPB, and any of the S$_1$, TT$_{pair}$, T$_{free}$ and the ground state. There is not yet a rigorous theory to describe these changes and, as mentioned before, including more states to the kinetic model increases the number of parameters, making the fit unreliable.

\begin{figure}

 \begin{center}
\includegraphics[width=.48\textwidth]{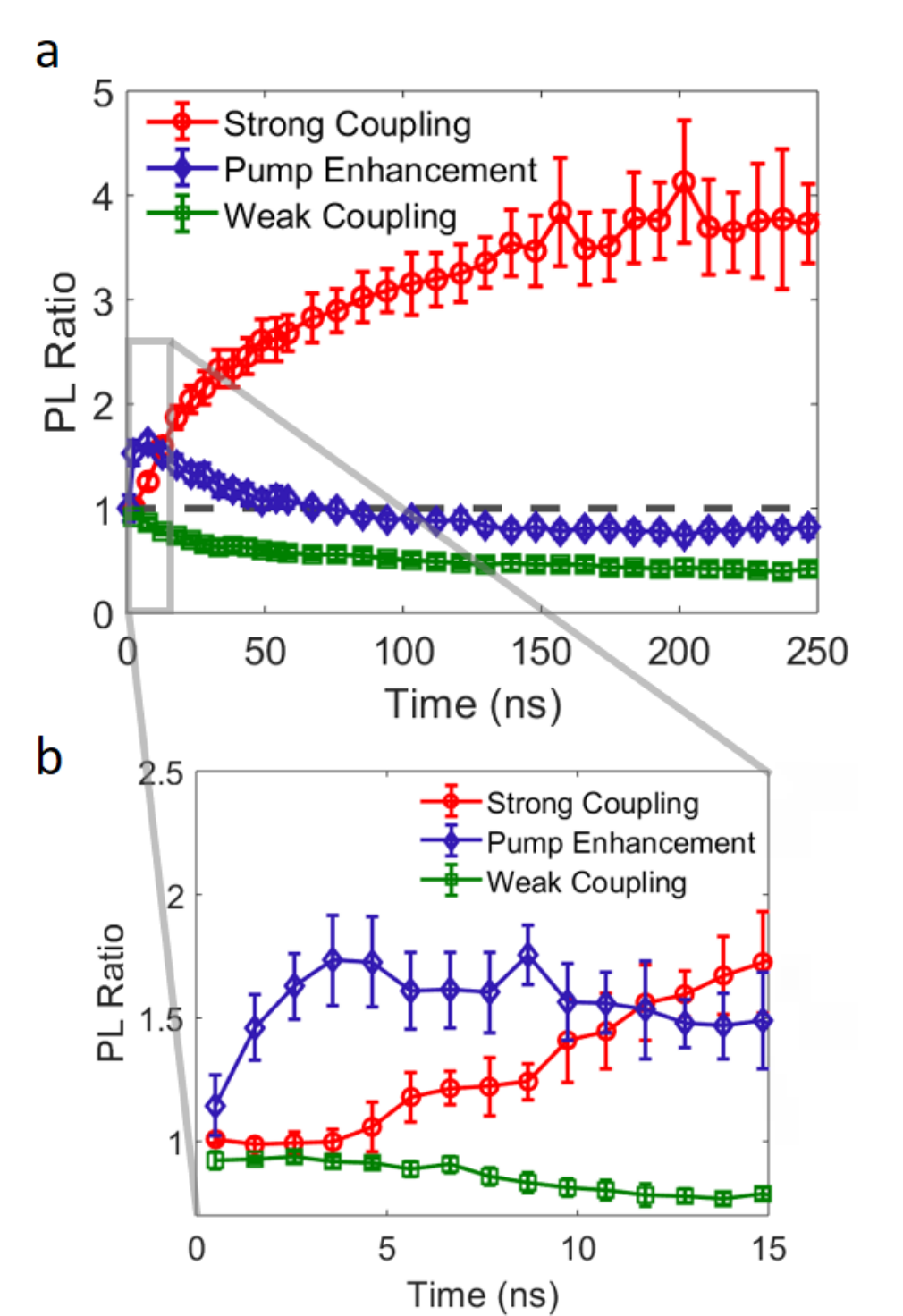}

\end{center}
  \caption{Ratios of fluorescence decay. \textbf{a} Ratio between fluorescence after normalization for the cases: on and outside the array in strong coupling (red circles), pump enhancement (blue diamonds) and on and outside the array in weak coupling (green squares). \textbf{b} Close up of the ratios during the first 15 ns. The ratio for the strong coupling case (red circles) stays close to one for the first 5 ns, and then increases rapidly.}
  
    \label{fgr:model}
\end{figure}

\section{Excited State Dynamics of Bare Tetracene at Different Excitation Intensities }

As the rate from free triplets to the triplet pair depends on the concentration of the triplets, increasing the excitation intensity should also affect the delayed fluorescence. To rule out the possibility that pump enhancement of the tetracene on the array, originating from scattering of the excitation beam with the nanoparticles, could cause the measured enhanced delayed fluorescence, we have done a control measurement where we experimentally reproduce the pump enhancement in bare tetracene by increasing the excitation intensity. From FDTD simulations, the pump enhancement was calculated to be a factor of 1.5 at 375 nm on the array (see Fig. S4 of the SI). Therefore, we compared two measurements of the TRPL where we increased the laser power by a factor of 1.5.

The ratio between the TRPL measured at high and low excitation intensities is plotted with the blue curve and diamonds in Figs. 5 (a) and (b). The decay curves are shown in SI (Figs. S5 and S7), where we also show that we can fit the effect of pump enhancement using the kinetic model by only changing the initial population of the singlet states (N$_{S}$). The fluorescence intensity is higher for high excitation intensities up to $\sim$50 ns (PL ratio larger than one in Figs. 5 (a) and (b)), as we can expect from the non-linear dependence of singlet fission on the triplet concentration. However, after the first few nanoseconds, the fluorescence decays faster due to the faster `depletion' of the triplet reservoir. We stress that the  differences between the effect of higher excitation intensity and strong coupling are very pronounced, as can be seen in the PL ratio of Figs. 5 (a) and (b). These differences rules out pump enhancement as the mechanism for the enhanced delayed fluorescence measured in the strongly coupled tetracene. 

\section{Excited State Dynamics of Weakly Coupled Tetracene Crystals}

As a second control experiment, we have designed a nanoparticle array with a periodicity of 200 $\times$ 420 nm$^2$ resulting in an SLR at 1.9 eV for $k_p=0$ rad/$\mu$m. This SLR which is at much lower energies than the exciton transition in tetracene. Similar extinction and emission measurements as those in Fig. 2, do not show an splitting into the LPB and UPB, and strong coupling for this sample (See Fig. S3 in the SI). The tetracene crystal is in this case weakly coupled to the nanoparticle array.

The PL ratio measured on the tetracene crystal on and outside the array is plotted with the green curve and open squares in Figs. 5 (a) and (b) (the normalized TRPL measurements of this sample are shown in Fig. S5 of the SI). This PL ratio shows a pronouncedly different behavior than the measurements on the strongly coupled sample: the contribution to the delayed fluorescence of the tetracene on the array is much lower than that of the tetracene outside the array. Using the kinetic model, we can reproduce this faster decay of tetracene weakly coupled to the array by considering a combination of pump enhancement, due to scattering of the excitation beam with the array, and a faster the decay of the S$_1$ state to the ground state (see SI Figs. S5 and S6 and table S4). This faster decay to the ground state corresponds to the Purcell enhancement due to the modified local density of optical states.\cite{Muskens2007} Note that the kinetic model can be used to explain these measurements as this sample is in the weakly coupling regime and no new states are formed.

Since the effect of weak coupling to the tetracene emission is very different from the strong coupling, we can exclude the metal as the origin of the enhanced delayed fluorescence of tetracene, which can be associated to strong coupling. For achieving the enhanced delayed PL, the resonance of the array should couple strongly to the tetracene.

\section{Conclusions}
We have demonstrated strong coupling of tetracene single crystals to collective plasmonic resonances (surface lattice resonances, SLRs) in open plasmonic cavities formed by arrays of silver nanoparticles. The strong coupling is evidenced by the splitting of the exciton energy in a lower polariton and an upper polariton band, with a Rabi energy of 210 meV. In the steady state emission, we have observed a wavelength dependent increase of the fluorescence, up to a factor of 7 at the energy of the lower polariton band. % The polariton hybrid states enable a new radiative decay channel for the long-lived triplet states,
Moreover, in time resolved PL (TRPL) measurements, we have observed an increase of the fractional contribution of the delayed fluorescence by almost a factor of four. With control measurements, we have shown that the enhanced delayed fluorescence measured in the strong coupling regime has a pronouncedly different behaviour than pump or Purcell enhancement. We have used a kinetic model to explain the TRPL of bare tetracene and the effects of pump and Purcell enhancement. However, the kinetic model was unable to explain the enhanced delayed fluorescence of tetracene in the strong coupling regime. 
From the analysis of the populations of different excited states in the tetracene crystal and the fact that the enhanced PL was on the same timescale as the free triplet lifetime, we suggest that long lived triplet states harvested by the lower polariton band may be at the origin of the enhanced delayed fluorescence. However, a more extensive theoretical model, including the complexity of the multilevel organic crystal coupled to to SLRs is required. The open structure defined by the nanoparticle array will facilitate the application of strong light-matter coupling in real devices, such as organic solar cells and light emitting diodes.

\section{Methods} \label{Methods}
\textbf{Silver nanoparticle array fabrication}. A layer of electron beam resist (ZEP520A) with a thickness of 120 nm was spin coated on top of a glass substrate. Conventional e-beam lithography (Raith EBPG 5250) was used to pattern the resist. After removal of the exposed resist, a 40 nm thick layer of silver was evaporated under $\sim 3 \times 10^{-8}$ Torr, and the remaining resist was removed. A 12 nm thick Si$_3$N$_4$ and an 8 nm thick SiO$_2$ protective layers were deposited to prevent oxidation of the nanoparticles. For better adhesion of the silver particles, a thin TiO$_2$ layer (2 nm) was deposited before evaporation of the silver.\\
\textbf{Tetracene crystals growth}. Tetracene crystals were fabricated by dissolving tetracene (98\% from Sigma Aldrich) in toluene until saturation. The solution was stirred on a hotplate at a temperature of 60$^\circ$C for 30 min inside a nitrogen glovebox. The solution was cooled down to 40$^\circ$C, filtered through a 0.22 $\mu$m PTFE filter (PerkinElmer), and dropcasted on the substrate. The thickness of the crystals was measured using a DektakXT profilometer at a load of 6 mg. However, as the stylus of the profilometer could damage the crystals, the specific thickness of the crystal used for the optical experiments was estimated by comparing its extinction to that of a crystal with a known thickness.\\
\textbf{Dispersion measurements with rotation stage.} The optical extinction (1-transmission) of the nanoparticle array covered with PS was measured with a collimated white light (Energetiq LDLS EQ-99) and with the sample mounted on a motorized rotation stage to change the angle of incidence. The extinction spectra were detected with an Ocean Optics USB 2000+ spectrometer. The measured spectra were referenced to the extinction measurements of tetracene outside the array and at the same angle. The large size of the arrays ($2.5 \times 2.5$ $\rm{mm}^2$) allows for incident angles over 50$^\circ$ with respect to the surface normal.\\
\textbf{Dispersion measurements with Fourier Microscope.} The extinction maps of the tetracene crystals were measured in a Fourier microscope. The rotation stage used for the extinction measurements of the bare array could not be used for these measurements as the diameter of the white light beam in that setup was larger than the tetracene crystal. The sample is illuminated through a 40x objective (Nikon CFI S Plan Fluor ELWD, NA 0.6) and collected with a 60x objective (Nikon CFI S Plan Fluor ELWD, NA 0.7). There was a 400 $\mu$m pinhole in the intermediate imaging plane such that light from an area of $\sim$7 $\mu$m diameter was measured in the Fourier plane. A spectrometer (Princeton Instruments SP2300) connected to a camera (Princeton Instruments ProEM:512) allows for mapping the dispersion as function of energy and angle. Extinction measurements of tetracene on the array were referenced to tetracene extinction outside the array. For measurements of the dispersion of the fluorescence from the samples, a 375 nm pulsed laser (LDH series Picoquant) was used for excitation.\\
\textbf{Confocal Microscope.} A Nikon confocal (C2) microscope with a CCD camera (DS-Fi2-U3) was used for imaging the sample. The images were taken by illuminating the sample with a X-Cite 120 white light source. For the fluorescence measurements, a 400 nm short pass filter is placed in the excitation beam path and a 447 nm long pass filter is used to collect the fluorescence image. Lifetime measurements of the excited states were done with a time-correlated single photon counting detector (TimeHarp 300 Picoquant) connected to the microscope and exciting the sample with a 375 nm pulsed laser operated at 2.5 MHz (Picoquant LDH series). The TRPL emission with a wavelength between 545 and 555 nm was collected using a band pass filter. The TRPL was also measured at the energy of the exciton with a band pass filter in the wavelength range 525 to 535 nm (measurements included in Fig. S8 of the SI).

\begin{acknowledgement}
We thank Reinder Coehoorn, Aart Ligthart and Wijnand Dijkstra for access to their chemical lab and assistance with fabrication techniques and we thank Erwin Zoethout for the XRD measurements.
We also acknowledge financial support from the Innovational Research Incentives Scheme of the Nederlandse Organisatie voor Wetenschappelijk Onderzoek (NWO) (Vici grant nr. 680-47-628) and to the NWO-Philips Industrial Partnership Program Nanophotonics for Solid State Lighting.\\

\end{acknowledgement}

\bibliography{Mendeley_Paper1}

\end{document}